# The Pricing of A Moving Barrier Option

## O, Hyong-chol

Centre of Basic Sciences, Kim Il Sung University, Pyongyang, D. P. R. of Korea,
Department of Applied Mathematics, Tong-ji University, Shanghai 200092, China

**Biography**: **Hyong-chol O**(1964-), Researcher (Corresponding Author: Tel: 021-65985699; Fax: 021-65982342,   E-mail: hyongchol_o@yahoo.com.cn)

**Abstract**: In this paper, we provided an analytical representation of the price of a barrier option with one type of special moving barrier. We consider the case that risk free rate, dividend rate and stock volatility are time dependent. We get a pricing formula and put call parity for barrier option when the moving barrier has a special relation with risk free rate, dividend rate and stock volatility.

## 1．Introduction

Barrier options are considered to be one of the types of path dependent options. The payoff of barrier options depend not only on the final price of the underlying asset but also on whether or not the underlying asset price has reached some barrier level during the life of the option. An **Out-barrier**(or **knock-out**) option is one that the option price becomes zero prior to expiration if the underlying asset price touches the **barrier**(or **out-strike price**). An **In-barrier**(or **knock-in**) option is one which only comes in existence if the asset price crosses the **in-barrier**. When the barrier is approached from below, the barrier option is called an **up-option**; otherwise, it is called a **down-option**. One can identify eight types of European barrier option, such as **down-and-out calls, Up-and-out calls, down-and-in puts, down-and-in puts**, etc. The barrier option with time dependent barrier value is called a **moving barrier option**[1].

The formula of various barrier options with constant risk free and dividend rates or volatility are considered in [1,4]. In particular they provided the price formula in the case that the barrier is exponential function of time. In [2, 3] are provided the price formula of barrier options using image solution method, but their image solution is no more a solution when risk free and dividend rates and volatility are not constant.

In general case, it is impossible to provide the price formula for moving barrier option even when risk free and dividend rate or volatility are constant[1, 255pp].

Here we provided the price formula for special class of the barrier functions when risk free and dividend rate or volatility depend on time.

Our results generalize the corresponding results of [1, 4] and the image solution of [3] is a special case of ours.

## 2. Moving Barrier Option Pricing Model

Consider an underlying asset(for example, a stock) whose price $S$ at time $t$ satisfies Ito stochastic differential equation. Let $r(t)$, $q(t)$ and $\sigma(t)$ be respectively risk free rate, dividend rate and stock volatility. Let $h(t)$ be a moving barrier. Then the price of *down-and-out call barrier option* $V(S,t)$ for $t < T$ satisfies the following (1)-(3).

$$\frac{\partial V}{\partial t} + \frac{\sigma^2(t)}{2} S^2 \frac{\partial^2 V}{\partial S^2} + (r(t)-q(t))S \frac{\partial V}{\partial S} - r(t)V = 0, \quad t>0,\ h(t)<S<\infty \quad (1)$$

$$V(S,T) = (S-K)^+ \quad (2)$$

$$V(h(t),t) = 0 \quad (3)$$

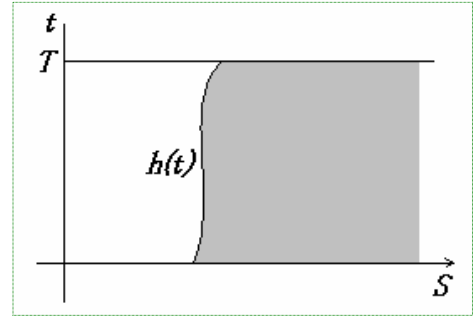

Here $K$ is strike price and $T$ is exercise date.

So this option stops its effect at the moment when the stock price touches the barrier from upper and otherwise, gives the final payoff of Eropean calls at the exercise date.

## 3. Analytical Formula of the Price of A Moving Barrier Option

Let

$$x = \ln(S/h(t)), \quad u(x(S),t) = V(S,t) \quad (4)$$

Then

$$\frac{\partial V}{\partial t} = \frac{\partial u}{\partial t} + \frac{\partial u}{\partial x}\left(-\frac{h'(t)}{h(t)}\right), \quad S\frac{\partial V}{\partial S} = \frac{\partial u}{\partial x}, \quad S^2\frac{\partial^2 V}{\partial S^2} = \frac{\partial^2 u}{\partial x^2} - \frac{\partial u}{\partial x}$$

and thus the terminal-boundary value problem (1)-(3) is transformed as follows.

$$\frac{\partial u}{\partial t} + \frac{\sigma^2(t)}{2}\frac{\partial^2 u}{\partial x^2} + \left(r(t)-q(t)-\frac{\sigma^2(t)}{2}-\frac{h'(t)}{h(t)}\right)\frac{\partial u}{\partial x} - r(t)u = 0, \quad t>0,\ 0<x<\infty \quad (5)$$

$$u(x,T) = (e^x h(T) - K)^+ \quad (6)$$

$$u(0,t) = 0 \quad (7)$$

Generally, it is difficult to provide analytic representation of solutions of the problem (5)-(7).
Let

$$u(x,t) = U(x,t) e^{a(t)x + b(t)} \quad (8)$$

Then

$$\frac{\partial u}{\partial t} = \frac{\partial U}{\partial t} e^{a(t)x+b(t)} + U e^{a(t)x+b(t)} (a'(t)x + b'(t)),$$

$$\frac{\partial u}{\partial x} = \frac{\partial U}{\partial x} e^{a(t)x+b(t)} + U e^{a(t)x+b(t)} a(t),$$

$$\frac{\partial^2 u}{\partial x^2} = \frac{\partial^2 U}{\partial x^2} e^{a(t)x+b(t)} + 2\frac{\partial U}{\partial x} e^{a(t)x+b(t)} a(t) + U e^{a(t)x+b(t)} a^2(t)$$

and thus, replacing upper expressions into (5) and eliminating $e^{a(t)x+b(t)}$, then

$$\frac{\partial U}{\partial t} + \frac{\sigma^2(t)}{2}\frac{\partial^2 U}{\partial x^2} + \left(r(t) - q(t) - \frac{\sigma^2(t)}{2} - \frac{h'(t)}{h(t)} + \sigma^2(t)a(t)\right)\frac{\partial u}{\partial x}$$

$$+ \left[a'(t)x + b'(t) + \frac{\sigma^2(t)a^2(t)}{2} + a(t)\left(r(t) - q(t) - \frac{\sigma^2(t)}{2} - \frac{h'(t)}{h(t)}\right) - r(t)\right]U = 0$$

Let

$$a(t) = -\sigma^{-2}(t)(r(t) - q(t) - \sigma^2(t)/2 - h'(t)/h(t))$$

$$b(t) = -\int_t^T \left(r(s) - \frac{\sigma^2(s)a^2(s)}{2} - a(s)\left(r(s) - q(s) - \frac{\sigma^2(s)}{2} - \frac{h'(s)}{h(s)}\right)\right)ds \quad (9)$$

$$= -\int_t^T \left(r(s) + \frac{\sigma^2(s)a^2(s)}{2}\right)ds$$

, then (5) is transformed into the following:

$$\frac{\partial U}{\partial t} + \frac{\sigma^2(t)}{2}\frac{\partial^2 U}{\partial x^2} + a'(t)xU = 0$$

By the time scale transformation of

$$\tau = \int_t^T \sigma^2(s)ds \quad (10)$$

the above equation is transformed into:

$$\frac{\partial U}{\partial \tau} - \frac{1}{2}\frac{\partial^2 U}{\partial x^2} + w(\tau)xU = 0, \quad 0 < \tau < T, x > 0 \quad (11)$$

where $w(\tau) = -a'(t)\sigma^{-2}(t)$

Since $b(T) = 0$, then under the transformation (8) and (10), the terminal and boundary conditions (6) and (7) are changed as follows:

$$U(x,0) = e^{-a(T)x}(e^x h(T) - K)^+ \quad (12)$$

$$U(0,\tau) = 0 \quad (13)$$

We can easily know that the equation (11) can be explicitly solved in the case when $a'(t) = 0$, that is, $a(t) \equiv \text{const.}$

Now, we explain the ***image solution method***. Consider the following problem:

$$\frac{\partial W}{\partial t} - \frac{1}{2}\frac{\partial^2 W}{\partial x^2} = 0, \quad t>0, x>0 \tag{14}$$

$$W(x,0) = f(x) \tag{15}$$

$$W(0,t) = 0 \tag{16}$$

Let

$$\bar{f}(x) = \begin{cases} f(x), & x>0 \\ 0, & x<0 \end{cases}$$

Let $W_1(x,t)$ be the solution of the following initial value problem:

$$\frac{\partial W}{\partial t} - \frac{1}{2}\frac{\partial^2 W}{\partial x^2} = 0, \quad t>0, -\infty<x<\infty$$

$$W(x,0) = \bar{f}(x)$$

Then by Poisson formula[5],

$$W_1(x,t) = \int_{-\infty}^{\infty} \frac{1}{\sqrt{2\pi t}} e^{-\frac{(x-\xi)^2}{2t}} \bar{f}(\xi)d\xi = \int_0^{\infty} \frac{1}{\sqrt{2\pi t}} e^{-\frac{(x-\xi)^2}{2t}} f(\xi)d\xi$$

Let $W_2(x,t) = W_1(-x,t)$, then $W_2(0,t) = W_1(0,t)$ and

$$W_2(x,t) = \int_0^{\infty} \frac{1}{\sqrt{2\pi t}} e^{-\frac{(x+\xi)^2}{2t}} f(\xi)d\xi$$

Then $W(x,t) = W_1(x,t) - W_2(x,t)$, $x>0$ is the solution of the problem (14)-(16) and is given by the following formula.

$$W(x,t) = \int_0^{\infty} \frac{1}{\sqrt{2\pi t}} \left( e^{-\frac{(x-\xi)^2}{2t}} - e^{-\frac{(x+\xi)^2}{2t}} \right) f(\xi)d\xi \tag{17}$$

Let *assume* that the *moving barrier function* ***h(t)*** has the following type:

$$h(t) = h(T)\exp\left(-\int_t^T (r(s) - q(s) + C\sigma^2(s))ds\right) \tag{18}$$

,here ***C*** is any constant. Then,

$$h'(t)/h(t) = r(t) - q(t) + C\sigma^2(t),$$

and thus in this case, (9) is changed as follows:

$$a(t) = C + 1/2$$
$$b(t) = -\int_t^T \left( r(s) + \frac{1}{2}\left(C + \frac{1}{2}\right)^2 \sigma^2(s) \right) ds, \quad b(T) = 0 \quad (19)$$

By transformation (8) and (10), the equation (5) is changed to

$$\frac{\partial U}{\partial \tau} - \frac{1}{2}\frac{\partial^2 U}{\partial x^2} = 0, \quad 0 < \tau < T, x > 0 \quad (20)$$

From the above considering, the solution of the initial-boundary value problem (20),(12) and (13) is :

$$U(x,\tau) = \int_0^\infty \frac{1}{\sqrt{2\pi\tau}} \left( e^{-\frac{(x-\xi)^2}{2\tau}} - e^{-\frac{(x+\xi)^2}{2\tau}} \right) e^{-a(T)\xi}(e^\xi h(T) - K)^+ d\xi \quad (21)$$

In what follows we use the following symbols for simplicity.

$$\bar{r} = \bar{r}(t,T) = \int_t^T r(s)ds, \quad \bar{q} = \bar{q}(t,T) = \int_t^T q(s)ds, \quad \overline{\sigma^2} = \overline{\sigma^2}(t,T) = \int_t^T \sigma^2(s)ds \quad (22)$$

From (18), (19) and (22), the following equalities are true:

$$C\overline{\sigma^2}(t,T) = -(\bar{r}(t,T) - \bar{q}(t,T)) - \ln\frac{h(t)}{h(T)}$$
$$b(t) = -\bar{r}(t,T) - \frac{1}{2}\left(C + \frac{1}{2}\right)^2 \overline{\sigma^2}(t,T) \quad (23)$$

In (22), $\tau = \overline{\sigma^2}(t,T)$, and so

$$u(x,t) = U(x,t)e^{a(t)x+b(t)} =$$
$$= e^{a(t)x+b(t)} \int_0^\infty \frac{1}{\sqrt{2\pi\overline{\sigma^2}}} \left( e^{-\frac{(x-\xi)^2}{2\overline{\sigma^2}}} - e^{-\frac{(x+\xi)^2}{2\overline{\sigma^2}}} \right) e^{-a(T)\xi}(e^\xi h(T) - K)^+ d\xi$$

$$= e^{a(t)x+b(t)} \int_0^\infty \frac{1}{\sqrt{2\pi\overline{\sigma^2}}} e^{-\frac{(x-\xi)^2}{2\overline{\sigma^2}}} e^{-a(T)\xi}(e^\xi h(T) - K)^+ d\xi$$
$$- e^{a(t)x+b(t)} \int_0^\infty \frac{1}{\sqrt{2\pi\overline{\sigma^2}}} e^{-\frac{(x+\xi)^2}{2\overline{\sigma^2}}} e^{-a(T)\xi}(e^\xi h(T) - K)^+ d\xi = I_1 + I_2 \quad (24)$$

Using (19) and (23), then

$$I_1 = e^{(C+\frac{1}{2})x+\overline{r}-\frac{1}{2}(C+\frac{1}{2})^2\overline{\sigma^2}} \int_0^\infty \frac{1}{\sqrt{2\pi\overline{\sigma^2}}} e^{-\frac{1}{2\overline{\sigma^2}}(x-\xi)^2} e^{-(C+\frac{1}{2})\xi} (e^\xi h(T) - K)^+ d\xi$$

$$= e^{-\overline{r}(t,T)} \int_{\ln\frac{K}{h(T)}}^\infty \frac{1}{\sqrt{2\pi\overline{\sigma^2}}} e^{-\frac{1}{2\overline{\sigma^2}}\left(x-\xi-\left(C+\frac{1}{2}\right)\overline{\sigma^2}\right)^2} (e^\xi h(T) - K)^+ d\xi$$

$$= e^{-\overline{r}(t,T)} \int_{\ln\frac{K}{h(T)}}^\infty \frac{h(T)}{\sqrt{2\pi\overline{\sigma^2}}} e^{-\frac{1}{2\overline{\sigma^2}}\left(x-\xi-\left(C+\frac{1}{2}\right)\overline{\sigma^2}\right)^2} e^\xi d\xi$$

$$- Ke^{-\overline{r}(t,T)} \int_{\ln\frac{K}{h(T)}}^\infty \frac{1}{\sqrt{2\pi\overline{\sigma^2}}} e^{-\frac{1}{2\overline{\sigma^2}}\left(x-\xi-\left(C+\frac{1}{2}\right)\overline{\sigma^2}\right)^2} d\xi = \Delta_1 + \Delta_2$$

$$-\Delta_2 = Ke^{-\overline{r}(t,T)} \int_{-\infty}^{\frac{x-\ln\frac{K}{h(T)}-\left(C+\frac{1}{2}\right)\overline{\sigma^2}}{\sqrt{\overline{\sigma^2}}}} \frac{1}{\sqrt{2\pi}} e^{-\frac{w^2}{2}} dw = Ke^{-\overline{r}(t,T)} N(d_1')$$

$$d_1' = \frac{x - \ln\frac{K}{h(T)} - \left(C+\frac{1}{2}\right)\overline{\sigma^2}(t,T)}{\sqrt{\overline{\sigma^2}(t,T)}}.$$

$$\Delta_1 = e^{-\overline{r}(t,T)} \int_{\ln\frac{K}{h(T)}}^\infty \frac{h(T)}{\sqrt{2\pi\overline{\sigma^2}}} e^{-\frac{1}{2\overline{\sigma^2}}\left(x-\xi-\left(C+\frac{1}{2}\right)\overline{\sigma^2}\right)^2 + \xi} d\xi$$

$$= e^x e^{-\overline{r}(t,T)-C\overline{\sigma^2}(t,T)} h(T) \int_{\ln\frac{K}{h(T)}}^\infty \frac{1}{\sqrt{2\pi\overline{\sigma^2}}} e^{-\frac{1}{2\overline{\sigma^2}}\left(x-\xi-\left(C-\frac{1}{2}\right)\overline{\sigma^2}\right)^2} d\xi$$

$$= e^x e^{-\overline{q}(t,T)} h(t) \int_{-\infty}^{\frac{x-\ln\frac{K}{h(T)}-\left(C-\frac{1}{2}\right)\overline{\sigma^2}}{\sqrt{\overline{\sigma^2}}}} \frac{1}{\sqrt{2\pi}} e^{-\frac{1}{2}w^2} dw = e^{-\overline{q}(t,T)} e^x h(t) N(d_1)$$

$$d_1 = \frac{x - \ln\frac{K}{h(T)} - \left(C-\frac{1}{2}\right)\overline{\sigma^2}(t,T)}{\sqrt{\overline{\sigma^2}(t,T)}}$$

$$I_2 = -e^{(C+\frac{1}{2})x+\overline{r}-\frac{1}{2}(C+\frac{1}{2})^2\overline{\sigma^2}} \int_0^\infty \frac{1}{\sqrt{2\pi\overline{\sigma^2}}} e^{-\frac{1}{2\overline{\sigma^2}}(x+\xi)^2} e^{-(C+\frac{1}{2})\xi} (e^\xi h(T) - K)^+ d\xi$$

$$= -e^{-\overline{r}(t,T)+2\left(C+\frac{1}{2}\right)x} \int_{\ln\frac{K}{h(T)}}^\infty \frac{1}{\sqrt{2\pi\overline{\sigma^2}}} e^{-\frac{1}{2\overline{\sigma^2}}\left(x+\xi+\left(C+\frac{1}{2}\right)\overline{\sigma^2}\right)^2} (e^\xi h(T) - K)^+ d\xi$$

$$= -e^{-\bar{r}(t,T)+2\left(C+\frac{1}{2}\right)x} \int_{\ln\frac{K}{h(T)}}^{\infty} \frac{h(T)}{\sqrt{2\pi\overline{\sigma^2}}} e^{-\frac{1}{2\overline{\sigma^2}}\left(x+\xi+\left(C+\frac{1}{2}\right)\overline{\sigma^2}\right)^2} e^{\xi} d\xi$$

$$+ Ke^{-\bar{r}(t,T)+2\left(C+\frac{1}{2}\right)x} \int_{\ln\frac{K}{h(T)}}^{\infty} \frac{1}{\sqrt{2\pi\overline{\sigma^2}}} e^{-\frac{1}{2\overline{\sigma^2}}\left(x+\xi+\left(C+\frac{1}{2}\right)\overline{\sigma^2}\right)^2} d\xi = \delta_1 + \delta_2$$

$$\delta_2 = Ke^{-\bar{r}(t,T)+2\left(C+\frac{1}{2}\right)x} \int_{\ln\frac{K}{h(T)}}^{\infty} \frac{1}{\sqrt{2\pi\overline{\sigma^2}}} e^{-\frac{1}{2\overline{\sigma^2}}\left(x+\xi+\left(C+\frac{1}{2}\right)\overline{\sigma^2}\right)^2} d\xi$$

$$= Ke^{-\bar{r}(t,T)}\left(e^x\right)^{2C+1} N(d_2')$$

$$d_2' = \frac{-x - \ln\frac{K}{h(T)} - \left(C+\frac{1}{2}\right)\overline{\sigma^2}(t,T)}{\sqrt{\overline{\sigma^2}(t,T)}}$$

$$-\delta_1 = e^{-\bar{r}(t,T)+2\left(C+\frac{1}{2}\right)x} \int_{\ln\frac{K}{h(T)}}^{\infty} \frac{h(T)}{\sqrt{2\pi\overline{\sigma^2}}} e^{-\frac{1}{2\overline{\sigma^2}}\left(x+\xi+\left(C+\frac{1}{2}\right)\overline{\sigma^2}\right)^2 + \xi} d\xi =$$

$$= e^{-\bar{r}(t,T)-C\overline{\sigma^2}(t,T)+2Cx} h(T) \int_{\ln\frac{K}{h(T)}}^{\infty} \frac{1}{\sqrt{2\pi\overline{\sigma^2}}} e^{-\frac{1}{2\overline{\sigma^2}}\left(x+\xi+\left(C-\frac{1}{2}\right)\overline{\sigma^2}\right)^2} d\xi$$

$$= e^{-\bar{q}(t,T)} h(t)\left(e^x\right)^{2C} \int_{-\infty}^{\frac{-x-\ln\frac{K}{h(T)}-\left(C-\frac{1}{2}\right)\overline{\sigma^2}}{\sqrt{\overline{\sigma^2}}}} \frac{1}{\sqrt{2\pi}} e^{-\frac{1}{2}w^2} dw = e^{-\bar{q}(t,T)} h(t)\left(e^x\right)^{2C} N(d_2)$$

$$d_2 = \frac{-x - \ln\frac{K}{h(T)} - \left(C-\frac{1}{2}\right)\overline{\sigma^2}(t,T)}{\sqrt{\overline{\sigma^2}(t,T)}}$$

So the solution of (5)-(7) is
$$u(x,t) = I_1 + I_2 = \Delta_1 + \Delta_2 + \delta_1 + \delta_2 =$$
$$= e^{-\bar{q}(t,T)} e^x h(t) N(d_1) - Ke^{-\bar{r}(t,T)} N(d_1')$$
$$- e^{-\bar{q}(t,T)} h(t)\left(e^x\right)^{2C} N(d_2) + Ke^{-\bar{r}(t,T)}\left(e^x\right)^{2C+1} N(d_2')$$

Considering (4) and using (23),

$$d_1 = \frac{\ln\frac{S}{K} + \bar{r}(t,T) - \bar{q}(t,T) + \frac{1}{2}\overline{\sigma^2}(t,T)}{\sqrt{\overline{\sigma^2}(t,T)}}$$

$$d_1' = \frac{\ln\frac{S}{K} + \overline{r}(t,T) - \overline{q}(t,T) - \frac{1}{2}\overline{\sigma^2}(t,T)}{\sqrt{\overline{\sigma^2}(t,T)}} = d_1 - \sqrt{\overline{\sigma^2}(t,T)}$$

$$d_2 = \frac{\ln\frac{h^2(t)}{SK} + \overline{r}(t,T) - \overline{q}(t,T) + \frac{1}{2}\overline{\sigma^2}(t,T)}{\sqrt{\overline{\sigma^2}(t,T)}}$$

$$d_2' = \frac{\ln\frac{h^2(t)}{SK} + \overline{r}(t,T) - \overline{q}(t,T) - \frac{1}{2}\overline{\sigma^2}(t,T)}{\sqrt{\overline{\sigma^2}(t,T)}} = d_2 - \sqrt{\overline{\sigma^2}(t,T)},$$

And so the price of *down-and-out call moving barrier option* is

$$V(S,t) = e^{-\overline{q}(t,T)}SN(d_1) - Ke^{-\overline{r}(t,T)}N(d_1')$$
$$- e^{-\overline{q}(t,T)}h(t)\left(\frac{S}{h(t)}\right)^{2C}N(d_2) + Ke^{-\overline{r}(t,T)}\left(\frac{S}{h(t)}\right)^{2C+1}N(d_2')$$
$$= V_{vanilla}(S,t,T,K) - \left(\frac{S}{h(t)}\right)^{2C+1}\left(\left(\frac{h^2(t)}{S}\right)e^{-\overline{q}(t,T)}N(d_2) - Ke^{-\overline{r}(t,T)}N(d_2')\right)$$
$$= V_{vanilla}(S,t,T,K) - \left(\frac{S}{h(t)}\right)^{2C+1}V_{vanilla}\left(\frac{h^2(t)}{S},t,T,K\right)$$

That is, the pricing formula is

$$V(S,t) = V_{vanilla}(S,t,T,K) - \left(\frac{S}{h(t)}\right)^{2C+1}V_{vanilla}\left(\frac{h^2(t)}{S},t,T,K\right) \quad (25)$$

Here C is the constant satisfying the following equality:

$$C = -(\overline{\sigma^2}(t,T))^{-1}\left(\overline{r}(t,T) - \overline{q}(t,T) + \ln\frac{h(t)}{h(T)}\right)$$

In this formula, the second term of (25)

$$V^*(S,t) = \left(\frac{S}{h(t)}\right)^{2C+1}\left(\left(\frac{h^2(t)}{S}\right)e^{-\overline{q}(t,T)}N(d_2) - Ke^{-\overline{r}(t,T)}N(d_2')\right)$$
$$= \left(\frac{S}{h(t)}\right)^{2C+1}V_{vanilla}\left(\frac{h^2(t)}{S},t,T,K\right) \quad (26)$$

is a solution of (1) such that

$$V^*(S, T) = 0, \quad S > h(t)$$
$$V^*(h(t), t) = V_{vanilla}(h(t), t), \quad t > 0.$$

$V^*(S, t)$ is called the *image solution* respect to the barrier $h(t)$. From out-in parity, $V^*(S, t)$

is the price of ***down-and-in call moving barrier option***.

**Theorem 1.** When the barrier function satisfies (18), then the prices of *down-and-out call moving barrier option* and *down-and-in call moving barrier option* are given by (25) and (26), respectively.

## 4. Put-Call Parity for Moving Barrier Option

In order to get the price of *down-and-out **put** moving barrier option*, let's consider the put-call parity for moving barrier options.

Let

$$V(S,t) = c_{down-and-out}(S,t) - p_{down-and-out}(S,t) \qquad (27)$$

Then $V$ satisfies (1) and (3), and furthermore

$$V(S,T) = S - K \qquad (28)$$

Using the above method, we solve the problem (1), (3), (28), then

$$\begin{aligned} V(S,t) &= e^{-\bar{q}(t,T)} S N(\overline{d_1}) - K e^{-\bar{r}(t,T)} N(\overline{d_1'}) \\ &\quad - e^{-\bar{q}(t,T)} h(t) \left(\frac{S}{h(t)}\right)^{2C} N(\overline{d_2}) + K e^{-\bar{r}(t,T)} \left(\frac{S}{h(t)}\right)^{2C+1} N(\overline{d_2'}) \\ &= e^{-\bar{q}(t,T)} S N(\overline{d_1}) - K e^{-\bar{r}(t,T)} N(\overline{d_1'}) \\ &\quad - \left(\frac{S}{h(t)}\right)^{2C+1} \left( e^{-\bar{q}(t,T)} \frac{h^2(t)}{S} N(\overline{d_2}) - K e^{-\bar{r}(t,T)} N(\overline{d_2'}) \right) \end{aligned} \qquad (29)$$

Here

$$\overline{d_1} = \frac{\ln \dfrac{S}{h(T)} + \bar{r}(t,T) - \bar{q}(t,T) + \dfrac{1}{2}\overline{\sigma^2}(t,T)}{\sqrt{\overline{\sigma^2}(t,T)}}$$

$$\overline{d_1'} = \frac{\ln \dfrac{S}{h(T)} + \bar{r}(t,T) - \bar{q}(t,T) - \dfrac{1}{2}\overline{\sigma^2}(t,T)}{\sqrt{\overline{\sigma^2}(t,T)}} = \overline{d_1} - \sqrt{\overline{\sigma^2}(t,T)}$$

$$\overline{d_2} = \frac{\ln\frac{h^2(t)}{Sh(T)} + \overline{r}(t,T) - \overline{q}(t,T) + \frac{1}{2}\overline{\sigma^2}(t,T)}{\sqrt{\overline{\sigma^2}(t,T)}}$$

$$\overline{d'_2} = \frac{\ln\frac{h^2(t)}{Sh(T)} + \overline{r}(t,T) - \overline{q}(t,T) - \frac{1}{2}\overline{\sigma^2}(t,T)}{\sqrt{\overline{\sigma^2}(t,T)}} = \overline{d_2} - \sqrt{\overline{\sigma^2}(t,T)}$$

So we can easily know that the first 2 terms represent the price of *vanilla call option with strike price h(T)*.

$$V(S,t) = V_{vanilla}(S,t,T,h(T)) - \left(\frac{S}{h(t)}\right)^{2C+1} V_{vanilla}\left(\frac{h^2(t)}{S},t,T,h(T)\right) \qquad (30)$$

Thus we proved the following theorem.

**Theorem 2**.( *Put-Call Parity for Moving Barrier Options*)

$$p_{down-and-out}(S,t) + e^{-\overline{q}(t,T)}SN(\overline{d_1}) = c_{down-and-out}(S,t) + Ke^{-\overline{r}(t,T)}N(\overline{d'_1})$$
$$+ \left(\frac{S}{h(t)}\right)^{2C+1}\left(e^{-\overline{q}(t,T)}\frac{h^2(t)}{S}N(\overline{d_2}) - Ke^{-\overline{r}(t,T)}N(\overline{d'_2})\right) \qquad (31)$$

In particular, if $r$, $q$ and $\sigma$ are all constants and $h(t) = S_B e^{-a(T-t)}$, then the put call parity formula is as follows:

$$p_{down-and-out}(S,t) + e^{-q(T-t)}SN(\hat{d}_1) = c_{down-and-out}(S,t) + Ke^{-r(T-t)}N(\hat{d}'_1)$$
$$+ e^{[1-\frac{2}{\sigma^2}(r-q-a)]a(T-t)}\left(\frac{S}{S_B}\right)^{1-\frac{2}{\sigma^2}(r-q-a)}\left(e^{-(q+2a)(T-t)}\frac{S_B^2}{S}N(\hat{d}_2) - Ke^{-r(T-t)}N(\hat{d}'_2)\right) \qquad (32)$$

Here

$$\hat{d}_1 = \frac{\ln\frac{S}{S_B} + (r-q+\frac{1}{2}\sigma^2)(T-t)}{\sigma\sqrt{(T-t)}}$$

$$\hat{d}'_1 = \hat{d}_1 - \sigma\sqrt{(T-t)} = \frac{\ln\frac{S}{S_B} + (r-q-\frac{1}{2}\sigma^2)(T-t)}{\sigma\sqrt{(T-t)}}$$

$$\hat{d}_2 = \frac{\ln\frac{S_B}{S} + (r-q-2a+\frac{1}{2}\sigma^2)(T-t)}{\sigma\sqrt{(T-t)}}$$

$$\hat{d}_2' = \hat{d}_2 - \sigma\sqrt{(T-t)} = \frac{\ln\frac{S_B}{S} + (r - q - 2a - \frac{1}{2}\sigma^2)(T-t)}{\sigma\sqrt{(T-t)}}$$

Using this relation and out-in parity, we can easily have the pricing formula of **down-and-out put** and **down-and-in put** moving barrier options.

Up-out or up-in moving barrier option pricing formula is provided by similar method, too.

## Acknowledgement

This is the first paper that the first author wrote in the area of mathematical finance as a *disciple* of Professor Li-shang Jiang. The author would like to thank to Professor Li-shang Jiang who led him to this theme, spent a lot of time for discussions and gave important keys to break through difficulties.

## References

[1] **Jiang, Lishang**(2003), Mathematical Modeling and Methods of Option Pricing, Higher Education Press, 247-260(in Chinese)
[2] **Buchen, P.**(2001), Image options and the road to barriers, Risk Mag., **14**, 127-130
[3] **Buchen, P.**(2004), The Pricing of dual-expiry exotics, Quantitative Finance,Vol.**4**, 101-108
[4] **Kwok, Y.K.**(1998), Mathematical models of Financial Derivatives, Springer-verlag, 246-254
[5] **Jiang, Lishang**(2003), Lectures on Equations of Mathematical Physics, Higher Education Press, 146-150(in Chinese)